\documentclass[12pt]{article}
\usepackage{amsfonts,amsmath,amssymb,mathrsfs}

\title{Simple Proof for Global Existence of Bohmian Trajectories}

\author{
Stefan Teufel\footnote{Mathematics Institute, University of
        Warwick, CV4 7AL Coventry, UK. E-mail:
        teufel@maths.warwick.ac.uk}
        \  and
Roderich Tumulka\footnote{Dipartimento di Fisica dell'Universit\`a
        di Genova and INFN sezione di Genova, Via Dodecaneso 33,
        16146 Genova, Italy.  E-mail:
        tumulka@mathematik.uni-muenchen.de}
}
\date{June 15, 2004}

\newcommand{\CCC}{\mathbb{C}} 
\newcommand{\RRR}{\mathbb{R}} 
\newcommand{\NNN}{\mathbb{N}} 
\newcommand{\EEE}{\mathbb{E}} 
\newcommand{\E}{\mathrm{e}} 
\newcommand{\I}{\mathrm{i}} 
\renewcommand{\Im}{\mathrm{Im}} 

\newcommand{\vx}{{\boldsymbol x}} 

\newcommand{\vz}{{\boldsymbol z}}

\newcommand{\vq}{{\boldsymbol q}}
\newcommand{\vQ}{{\boldsymbol Q}}
\newcommand{\vA}{{\boldsymbol A}}
\newcommand{\vB}{{\boldsymbol B}}
\newcommand{\valpha}{{\boldsymbol \alpha}}
\newcommand{\vsigma}{{\boldsymbol \sigma}}

\newcommand{\sgamma}{\Gamma}
\newcommand{\sj}{J}

\newcommand{\conf}{\mathcal{Q}} 
\renewcommand{\div}{\mathrm{div}\,} 
\newcommand{\flow}{\varphi}

\newcommand{\cemetery}{\diamondsuit} 
\newcommand{\Vari}{L} %
\newcommand{\Bdd}{\mathscr{B}} 
\newcommand{\Borel}{\mathscr{A}} 
\newcommand{\Nodes}{\mathscr{N}} 
\newcommand{\flowim}{\mathscr{I}} 
\newcommand{\measure}{\mu} 
\newcommand{\Sing}{S} 

\newtheorem{theorem}{Theorem}
\newtheorem{corollary}{Corollary}
\newtheorem{lemma}{Lemma}
\newenvironment{proof}[1][]{\noindent \textit{Proof#1.}\ }
  {\hfill$\square$\bigskip }

\begin{document}\maketitle
\begin{abstract}
  We address the question whether Bohmian trajectories exist for all
  times. Bohmian trajectories are solutions of an ordinary
  differential equation involving  a wavefunction obeying either
  the Schr\"odinger or the Dirac equation. Some trajectories may end
  in finite time, for example by running into a node of the
  wavefunction, where the law of motion is ill-defined. The aim is to
  show, under suitable assumptions on the initial wavefunction and the
  potential, global existence of almost all solutions.   We provide
   a simpler and more transparent proof of the known
  global existence result for spinless Schr\"odinger particles and
  extend the result to particles with spin, to the presence of
  magnetic fields, and to Dirac wavefunctions. Our main new result are  conditions on the current vector field on
  configuration-space-time
  which are sufficient for almost-sure global existence.

  \medskip

\end{abstract}

\section{Introduction}

We study  a mathematical question arising from and relevant to
Bohmian mechanics \cite{Bohm52, BellHidden, qe, survey, DetlefBuch}
and its variant based on the Dirac equation \cite{Bohm53, BohmHiley}
(henceforth referred to as the ``Bohm--Dirac theory'').  In these
theories, the motion of particles is defined by ordinary
differential equations (ODEs) involving the wavefunction, see
\eqref{Bohm} and \eqref{BohmDirac} below.  The mathematical question
we address is global existence, i.e., whether (under what conditions
and how often) the particle trajectories are well defined for all
times. One obstruction to global existence is that the velocity
given by \eqref{Bohm} or \eqref{BohmDirac} is singular at the nodes
(i.e., zeros) of the wavefunction. In particular, there are
trajectories that are not defined for all times because they run
into a node. Thus, the strongest statement one can expect to be true
is that global existence holds for \emph{almost all} solutions of
the equation of motion. As we show, this is in fact true for
suitable potentials and initial wavefunctions. As a by-product, one
obtains from almost-sure global existence the equivariance of the
$|\psi_t|^2$ distributions.

The relevance of Bohmian mechanics to the foundations of quantum
mechanics arises from the fact that a world governed by Bohmian
mechanics satisfies all probability rules of quantum mechanics
\cite{Bohm52, BellHidden, qe, survey, DetlefBuch}. Bohmian mechanics
thus provides an example of a ``quantum theory without observers,''
one in which no reference to observers is needed for the formulation
of the theory, and an explanation of the quantum probabilities in
terms of objective events.

The authoritative paper on global existence of Bohmian trajectories
is by Berndl et al.~\cite{bmex}; see also \cite{Karinthesis}. We
note that the proof given by Holland~\cite[p.~85]{Holland} is
incorrect (see \cite{bmex} for details). We also remark that the
general existence theory for first order ODEs with velocity  vector
fields that are not Lipschitz but only in some Sobolev space
\cite{DiPernaLions} does not apply to Bohmian trajectories. The
results of \cite{DiPernaLions} hold for vector fields with bounded
divergence, while  the divergence of  a Bohmian velocity field, such
as in \eqref{Bohm} and \eqref{Dirac}, typically diverges at nodes of
the wave function.
 Berndl et al.~\cite{bmex}
already proved almost-sure global existence for suitable potentials
and initial wavefunctions; while they give a proof only for spinless
nonrelativistic particles, a similar proof could presumably be
devised for Bohmian mechanics with spin \cite{BellHidden, survey}
and the Bohm--Dirac theory.  We provide here an alternative proof
that is shorter and more transparent than the proof by Berndl et al.
Our result covers all cases covered by their existence theorem; in
addition, our result also covers Bohmian mechanics with spin and
magnetic fields and Bohm--Dirac theory; for the latter our result
and its proof become particularly simple thanks to the fact that the
Bohm--Dirac velocities are bounded by the speed of light. Even more
generally, our result can be applied to any Bohm-type dynamics, as
we formulate conditions on the current vector field on
configuration-space-time that are sufficient for almost-sure global
existence.

There are three ways in which a trajectory can fail to exist
globally: it can approach a node of the wavefunction (where the
equation of motion is not defined), it can approach a singularity of
the potential (where the equation of motion need not be defined), or
it can escape to infinity in finite time. Hence, the main work of
any existence proof for Bohm-type dynamics is to show that almost
every trajectory avoids the ``bad points'' (nodes, singularities,
infinity) in configuration space.  The method of Berndl et al.\ is
based on estimating the probability flux across surfaces surrounding
the bad points and pushing these surfaces closer to the bad points;
in the limit in which the surfaces reach the bad points, the flux
vanishes.

The advantage of our approach is that it does not require skillful
estimates and does not involve limits. Instead, our method is based
on considering a suitable nonnegative quantity along the trajectory
that becomes infinite when the trajectory approaches a bad point; if
such a quantity has finite expectation, at least locally, then the
set of initial configurations for which it becomes infinite must be
a null set. That the expectation be locally finite can be
paraphrased as an integral condition on the  current vector field.

To illustrate our method, we briefly describe an argument of this
kind: the total distance $D$ traveled in configuration space in the
time interval $[0,T]$ becomes infinite when the trajectory escapes
to infinity during $[0,T]$. To prove that $D$ is almost surely
finite, we prove that it has finite expectation. A calculation shows
that
\begin{equation}\label{ED}
\EEE D\leq \int\limits_0^T
\hspace{-2pt}dt\,\int\limits_{\RRR^{3N}}\hspace{-3pt}dq\, |\sj|\,,
\end{equation}
where $\sj$ is the spatial component of the current vector field.
Thus, the finiteness of the right hand side of \eqref{ED} is a
natural condition on the current ensuring that almost no trajectory
escapes to infinity in $[0,T]$.

This argument was already sketched in \cite{schwerpunkt}; it was
inspired by a similar consideration in the global existence proof of
\cite{crex} for Bell's jump process for lattice quantum field
theory, another Markov process depending on a wavefunction $\psi_t$
and having distribution $|\psi_t|^2$ at any time $t$; the
 quantity considered there was the number of jumps during
$[0,T]$. Finally, a related argument was also described in
Remark~3.4.6 of \cite{bmex}, see our Remark~5 for a comparison.

Our method is also in a way more elementary than that of Berndl et
al.: we do not make use of the nontrivial fact that the
$|\psi|^2$-probability of crossing a surface $\Sigma$ in
configuration-space-time  is bounded by $\int_\Sigma |d\sigma \cdot
j|$ where $d\sigma$ is the normal on $\Sigma$ with length equal to
the area of the surface element and $j$ is the current vector field.
Indeed, we use this fact only for surfaces lying in $t=$ const.\
slices of configuration-space-time, for which it is much simpler to
prove, see Lemma~\ref{Lrhoj0}. To be sure, the   statement about
general surfaces is interesting in its own right and also relevant
to other applications such as scattering theory, but its proof takes
several pages in \cite{Karinthesis}.

While our innovation concerns sufficient conditions on the current
for almost-sure global existence, there remains the functional
analytic question of deriving these conditions from conditions on
the potential and the initial wavefunction. We carry this out in
several example cases but contribute nothing original; we employ the
same arguments as Berndl et al.\ or standard arguments.

This article is organized as follows. In Section~\ref{sec:setup} we
give the definition of Bohmian trajectories for both the
Schr\"odinger and the Dirac equation; we elucidate the relevance of
the  current vector field to the trajectories and their
distribution. In Section~\ref{sec:current} we state and prove our
results in terms of a current vector field. In
Section~\ref{sec:Dirac} we state and prove our results for
Bohm--Dirac theory. Finally, in Section~\ref{sec:Schr} we state and
prove our results for Bohmian mechanics.

\section{Setup}\label{sec:setup}

We briefly recall Bohmian mechanics and the Bohm--Dirac theory for a
system of $N$ particles. Then we describe what singularities we will
allow in the potential.  Finally, we point out how for both Bohmian
mechanics and Bohm--Dirac theory the trajectories arise from  a
current vector field on configuration-space-time.

\subsection{Equations of Motion}

In Bohmian mechanics, the wavefunction is a function $\psi: \RRR
\times \RRR^{3N} \to \CCC^k$ where $\RRR$ represents the time axis,
$\RRR^{3N}$ the configuration space of $N$ particles, and $\CCC^k$
the value space of the wavefunction representing the internal
degrees of freedom of the particles such as spin (and possibly quark
flavor etc.). $\psi = \psi(t, \vq_1, \ldots, \vq_N)$ evolves
according to the Schr\"odinger equation
\begin{equation}\label{Schr}
  \I \hbar \frac{\partial \psi}{\partial t} = - \sum_{i=1}^N
  \frac{\hbar^2}{2m_i} \bigl( \nabla_{\vq_i} - \tfrac{\I e_i}{c\hbar} \vA
  (\vq_i) \bigr)^2 \psi + V(\vq_1, \ldots, \vq_N) \psi,
\end{equation}
where $m_i$ and $e_i$ denote mass and charge of the $i$-th particle,
$c$ the speed of light, $\vA$ is the external electromagnetic vector
potential, and $V$ is the potential, which may be Hermitian $k
\times k$-matrix valued. For particles with spin in the presence of
magnetic fields, the potential includes a term $\sum_i \frac{\hbar
e_i}{2m_ic}(\nabla \times \vA) (\vq_i) \cdot \vsigma_i$ where
$\vsigma_i$ is the vector of spin operators (Pauli matrices for
spin-$\tfrac{1}{2}$) acting on the spin index of particle $i$; this
form of the Schr\"odinger equation is known as the Pauli equation.

The law of motion for the trajectory $\vQ_i(t)$ of the $i$-th
particle reads
\begin{equation}\label{Bohm}
  \frac{d \vQ_i}{dt}(t) = \frac{\hbar}{m_i} \Im \frac{\psi^* \, \bigl(
  \nabla_{\vq_i} - \tfrac{\I e_i}{c\hbar} \vA (\vq_i) \bigr) \psi} {\psi^* \,
  \psi} (t,Q(t)),
\end{equation}
where $Q = (\vQ_1, \ldots, \vQ_N)$ is the configuration, and $\psi^*
\phi$ denotes the inner product in $\CCC^k$. The right hand side of
\eqref{Bohm} is ill-defined when and only when either $\psi(t,Q)=0$
(node of $\psi$) or $\psi$ is not differentiable at $(t,Q)$. For an
explicit example of a trajectory that runs into a node of $\psi$,
see \cite{bmex}.

In Bohm--Dirac theory, the wavefunction is a function $\psi: \RRR
\times \RRR^{3N} \to \CCC^{4^N} = (\CCC^4)^{\otimes N}$ evolving
according to the Dirac equation
\begin{equation}\label{Dirac}
  \I \hbar \frac{\partial \psi}{\partial t} = - \sum_{i=1}^N \I c
  \hbar \valpha_i \cdot \nabla_{\vq_i} \psi + V(\vq_1, \ldots, \vq_N) \psi,
\end{equation}
where   $\valpha_i$ denotes the vector of Dirac alpha matrices
acting on the spin index of particle $i$; we have included the mass
terms in the potential $V$, which is Hermitian $4^N \times
4^N$-matrix valued. In the presence of magnetic fields, $V$ includes
a term $-\sum_i  e_i \vA(\vq_i) \cdot \valpha_i$.

The law of motion for the trajectory $\vQ_i(t)$ of the $i$-th
particle reads
\begin{equation}\label{BohmDirac}
  \frac{d \vQ_i}{dt}(t) = c \frac{\psi^* \, \valpha_i \, \psi}
  {\psi^* \, \psi} (t,Q(t))\,.
\end{equation}
 The right hand side is
ill-defined at nodes of $\psi$ and only there.

\subsection{Singularities of the Potential}

Among the physically relevant examples of potentials $V=V(\vq_1,
\ldots, \vq_N)$ is the Coulomb potential,
\begin{equation}\label{Coulomb}
  V(\vq_1, \ldots, \vq_N) = \sum_{i<j} \frac{e_i e_j}{|\vq_i -
  \vq_j|}\,
\end{equation}
which is singular at coincidence configurations (those with $\vq_i =
\vq_j$ for some $i \neq j$).  This motivates us to allow that $V$ is
defined only on a subset $\conf\subset \RRR^{3N}$; e.g.\ in the case
of Coulomb interaction, $\conf$ is the set of non-coincidence
configurations; in the case of an external Coulomb potential
generated by charges located at $\vz_1,\ldots,\vz_M$,
$\conf=(\RRR^3\setminus\{\vz_1,\ldots,\vz_M\})^N$. One cannot expect
a Schr\"odinger wavefunction to be differentiable on the singular
set $\RRR^{3N}\setminus\conf$ of the potential, as exemplified by
the ground state of the hydrogen atom, which is proportional to
$\exp(-\lambda |\vq|)$ for suitable $\lambda>0$. Thus, the right
hand side of \eqref{Bohm} may be ill-defined on
$\RRR^{3N}\setminus\conf$, and we will use differentiability of
$\psi$ only on $\conf$. For the Coulomb interaction and the external
Coulomb potential, $\conf$ is of the form $\conf=\RRR^d\setminus
\cup_{\ell=1}^m\Sing_\ell$, where $\Sing_\ell$ are hyperplanes. Our
method of proof allows somewhat weaker assumptions:

A closed set $\Sing\subset\RRR^d$ is admissible, if there is a
$\delta>0$ such that the distance function $q\mapsto {\rm
dist}(q,\Sing)$ is differentiable on the open set
$(\Sing+\delta)\setminus \Sing$, where $\Sing+\delta = \{q\in\RRR^d:
{\rm dist}(q,\Sing)<\delta\}$. Then the configuration space $\conf$
is
\begin{equation}\label{conf}
  \text{either } \conf = \RRR^d \text{ or } \conf = \RRR^d \setminus
  \bigcup_{\ell =1}^m \Sing_\ell,
\end{equation}
where $\Sing_1, \ldots, \Sing_m$ are admissible sets.
 For example
hyperplanes are obviously admissible sets.

\subsection{The Current Vector Field}\label{sec:CVF}

There is a common structure behind the laws of motion \eqref{Bohm}
and \eqref{BohmDirac}: they are of the form
\begin{equation}\label{dQdtj}
  \frac{dQ}{dt}(t) = \frac{\sj(t,Q(t))}{j^0(t,Q(t))}
\end{equation}
where $j=(j^0, \sj)$ is the current vector field on
configuration-space-time $\RRR \times \conf$, defined by
\begin{equation}\label{Bohmj}
  j = \Bigl( |\psi|^2, \tfrac{\hbar}{m_1} \Im \, \psi^* \bigl(
  \nabla_{\vq_1} - \tfrac{\I e_1}{c\hbar} \vA (\vq_1) \bigr) \psi,
  \ldots, \tfrac{\hbar}{m_N} \Im \, \psi^* \bigl(
  \nabla_{\vq_N} - \tfrac{\I e_N}{c\hbar} \vA (\vq_N) \bigr) \psi \Bigr)
\end{equation}
in the Schr\"odinger case and
\begin{equation}\label{BohmDiracj}
  j = \Bigl( |\psi|^2, c \, \psi^* \valpha_1 \psi, \ldots, c \,
  \psi^* \valpha_N \psi \Bigr)
\end{equation}
in the Dirac case. Provided that $\psi$ is sufficiently
differentiable, $j$ has the following properties, which we take to
be the defining properties of a   \emph{current vector field}:
\begin{subequations}\label{current}
\begin{align}\label{10a}
  &j =(j^0,\sj) \text{ is a $C^1$ vector field on } \RRR \times
  \conf\\\label{10b} & \div j = \sum_{\mu = 0}^d \partial_\mu j^\mu = 0\\ \label{10c}&j^0 >
  0 \text{ whenever } j \neq 0 \\[4pt] & \hspace{-2pt}\int\limits_{\conf} dq \: j^0(t,q)
  = 1 \quad \forall t \in \RRR \,. \label{j0norm}
\end{align}
\end{subequations}
We will call points in $\Nodes=\{(t,q)\in\RRR \times\conf:
j(t,q)=0\}$ the \emph{nodes of} $j$. We write $\Nodes_t=\{q\in
\conf: j(t,q)=0\}$ for the set of nodes at time $t$.

Let $Q_q(t)$ denote the maximal solution of \eqref{dQdtj} starting
in $q\in\conf\setminus\Nodes_0$ defined for
$t\in(\tau_q^-,\tau_q^+)$.
 It is a
reparameterization of an integral curve of $j$, see
Remark~\ref{remark8} for details. We will formulate our existence
theorem first purely in terms of the current vector field, and then
apply our result to the currents \eqref{Bohmj} and
\eqref{BohmDiracj}.

\subsection{Equivariance}\label{sec:equi}

We now explain the notion of equivariance, and what needs to be
shown to prove equivariance. We first remark that equivariance is a
crucial property of Bohm-type dynamics, in fact the basis of the
statistical analysis of Bohmian mechanics \cite{qe} and thus the
basis of the agreement between the predictions of Bohmian mechanics
and the prescriptions of quantum mechanics. We also remark that,
while full equivariance will be a consequence of the existence
result, a kind of partial equivariance can be obtained before, see
Lemma~\ref{Lrhoj0} below; our existence proof will exploit this
partial equivariance.

Before we define equivariance, we introduce some notation.  Let
$\Borel (\conf)$ denote the Borel $\sigma$-algebra of $\conf$.  Let
$\measure_t$ be the measure on $\Borel (\conf)$ with density
$j^0(t)$ relative to Lebesgue measure,
\begin{equation}\label{measuredef}
  \measure_t(B) = \int\limits_B dq \: j^0(t,q)
\end{equation}
for all $B \in \Borel (\conf)$. By \eqref{j0norm}, $\measure_t$ is a
probability measure; in Bohmian mechanics and Bohm--Dirac theory,
$\measure_t$ is the $|\psi(t)|^2$ distribution.  We introduce a
formal cemetery configuration $\cemetery$ and   set $Q_q(t)
:=\cemetery$ for all $t\notin (\tau_q^-,\tau_q^+)$, respectively, if
$(0,q)$ is a node of $j$, $Q_q(t):=\cemetery$ for all $t\not=0$.
 Let $\flow_t: \conf \to \conf
\cup \{\cemetery\}$, $\flow_t(q) = Q_q(t)$, denote the flow map of
\eqref{dQdtj}, and let $\flow: \RRR \times \conf \to \RRR \times
(\conf \cup \{ \cemetery \})$ be the flow map on
configuration-space-time defined by $\flow(t,q) = (t,
\flow_{t}(q))$.  Let $\conf_t = \{q \in \conf\setminus\Nodes_0:
\tau_q^-<
 t<\tau_q^+ \} = \flow_t^{-1}(\conf)$. Standard theorems (see, e.g.,
Chapter~II of \cite{ode}) on ODEs imply that $\flow$ is $C^1$ on the
maximal domain $\{(t,q) \in \RRR \times (\conf\setminus\Nodes_0):
\flow_t(q) \neq \cemetery \}$, which is open; in particular, also
$\conf_t$ is an open set.

Let $\rho_t$ be the distribution of $Q_q(t)$ if $q$ has distribution
$\measure_0$, i.e.,
\begin{equation}\label{rhodef}
  \rho_t = \measure_0 \circ \flow_t^{-1} \,.
\end{equation}

One says that \emph{the family of measures $\measure_t$ is
equivariant on the time interval $I$} if $\rho_t = \measure_t$ for
all $t \in I$. (The interval $I$ may be finite or infinite.)

Let $\flowim_t := \flow_t(\conf_t)=\flow_t (\conf) \cap \conf$ be
the image of the flow map in $\conf$ at time $t$. The following
lemma formulates ``partial equivariance.''

\begin{lemma}\label{Lrhoj0}
Let   $j=(j^0,\sj)$ satisfy \eqref{10a}, \eqref{10b}, and
\eqref{10c}.
  Then for all $B \in \Borel(\conf)$ and all $t \in \RRR$,
  \begin{equation}\label{rhomeasure}
    \rho_t(B) = \measure_t (B\cap\flowim_t) \,.
  \end{equation}
\end{lemma}

We know of two ways of proving this lemma, requiring comparable
effort. One proof, given in \cite{bmex} and in more detail in
\cite{Karinthesis}, goes as follows. $\rho_t$ has a density that
obeys a continuity equation, and $j^0$ satisfies the same continuity
equation. By uniqueness of solutions of this linear partial
differential equation, one obtains that $j^0(t)$ coincides with the
density of $\rho_t$ on $\flowim_t$. An alternative proof, which we
give below, is based on applying the Ostrogradski--Gauss integral
formula to $j$ on a cylinder formed by the trajectories over a
polyhedron in $\conf$.

\medskip
\newcommand{\chain}{F}
\newcommand{\schain}{E}
\begin{proof}[ of Lemma~\ref{Lrhoj0}]
Without loss of generality, $t>0$. For any $d$-chain of singular
simplices $\schain$ in $\conf_t$, the cylinder $\chain$ formed by
the trajectories over $\schain$, $\chain = \flow ([0,t] \times
\schain)$, is a $d+1$-chain in configuration-space-time $\RRR \times
\conf$.  Applying the Ostrogradski--Gauss integral formula to $j$
and $\chain$, we obtain
\[
  0\stackrel{\eqref{10b}}{=} \int\limits_\chain dt \, dq \, \div j = \int\limits_{\partial \chain}
  d\sigma \cdot j
\]
where $d\sigma$ is the outward pointing surface normal with length
$|d\sigma|$ equal to the area of the surface element. The surface
$\partial \chain$ of the cylinder consists of three parts: the
mantle $\flow([0,t] \times \partial \schain)$, the lid $\flow(\{t\}
\times \schain )$, and the bottom $\flow(\{0\} \times \schain )$.
The integral over the mantle vanishes as the mantle consists of
integral curves of $j$ and is thus tangent to $j$. The integral over
the lid is $\int_{\flow_t(\schain)} dq \: j^0(t,q)$ and that over
the bottom is $- \int_\schain dq\: j^0(0,q)$. Therefore, we obtain
\[
  0 = \measure_t (\flow_t(\schain)) - \rho_0(\schain) = \measure_t (\flow_t(\schain)) - \rho_t(\flow_t(\schain)) \,.
\]

Any two measures that agree on the $d$-chains (and thus in
particular on the compact rectangles) agree on a $\cap$-stable
generator of the $\sigma$-algebra $\Borel (\conf_t)$ and are, by a
standard theorem, equal. Since $\flow_t$ is a bijection $\conf_t \to
\flowim_t$, we obtain \eqref{rhomeasure}.
\end{proof}

What remains to be shown to prove equivariance is that $\measure_t
(\conf \setminus \flowim_t) =0$.

\section{A General Existence Theorem}\label{sec:current}

Let $\Bdd(\conf)$ denote the set of all bounded  Borel sets in
$\conf$.

\begin{theorem}\label{thm1} Let $\conf\subset\RRR^d$ be a configuration space as
defined in \eqref{conf} and let   $j=(j^0,\sj)$ be a current as
defined in \eqref{current}. Let $T>0$ and let $\flow_t:
  \conf \to \conf \cup \{\cemetery\}$ denote the flow map of
 \eqref{dQdtj}.
  Suppose that
  \begin{equation}\label{ass1}
  \forall B \in
    \Bdd(\conf):\quad
    \int\limits_0^T dt \int\limits_{\flow_t(B) \setminus
    \{\cemetery\} } \hspace{-1.2em} dq \:\:\:
    \left| \left( \frac{\partial}{\partial t} + \frac{\sj}{j^0} \cdot
  \nabla_q\right) j^0(t,q) \right| < \infty \,,
  \end{equation}
  \begin{equation}\label{ass3}
  \forall B \in
    \Bdd(\conf):\quad
    \int\limits_0^T dt \int\limits_{\flow_t(B) \setminus
     \{\cemetery\}
    } \hspace{-1.2em} dq \:\:\: \left| \sj (t,q)\cdot\frac{q}{|q|}\right| < \infty
    \,,
  \end{equation}
and, if $\conf=\RRR^d\setminus \cup_\ell \Sing_\ell$,  in addition
that for every $\ell\in\{1,\ldots,m\}$,
\begin{equation}\label{ass4}
\exists \delta>0\,\,\, \forall B \in
    \Bdd(\conf) :\quad
\int\limits_0^T dt \int\limits_{\flow_t(B) \setminus \{\cemetery\}
    }\hspace{-1.2em} dq \,{\bf 1}\big(q\in (\Sing_\ell+\delta)\big)\, \frac{\left| \sj(t,q)\cdot e_\ell(q)\right|}{{\rm
    dist}(q,\Sing_\ell)}<\infty\, .
\end{equation}
Here  ${\rm dist}(q,\Sing_\ell)$ is the Euclidean distance of $q$
from $\Sing_\ell$ and $e_\ell(q)=- \nabla_q{\rm dist}(q,\Sing_\ell)$
is the radial unit vector towards $\Sing_\ell$ at $q\in\conf$.
Recall that for $\delta$ sufficiently small the   distance function
is differentiable on $\Sing_\ell+\delta$.
\smallskip

  Then for almost every $q \in \conf$ relative to the measure
  $\measure_0(dq)  = j^0(0,q) \, dq$, the solution of \eqref{dQdtj} starting at $Q(0)=q$
  exists at least up to time $T$, and  the family of measures
  $\measure_t(dq) =j^0(t,q) \, dq$  is equivariant on $[0,T]$.  In particular, if
  \eqref{ass1},
  \eqref{ass3} and, if appropriate, \eqref{ass4} are true for every $T>0$, then
  for $\measure_0$-almost every $q \in \conf$ the solution of \eqref{dQdtj} starting at $q$ exists for all
  times $t\geq 0$.
\end{theorem}

\noindent {\bf Remarks:}
\begin{enumerate}
\item  We can formulate the meaning of each of the conditions
\eqref{ass1}, \eqref{ass3}, and \eqref{ass4} as follows. If
\eqref{ass1} holds, then $\mu_0$-almost no trajectory approaches a
node during $[0,T]$. If \eqref{ass3} holds, then $\mu_0$-almost no
trajectory escapes to $\infty$ during $[0,T]$. If \eqref{ass4}
holds, then $\mu_0$-almost no trajectory approaches a point in the
singular set $\cup_{\ell=1}^m \Sing_\ell$ during $[0,T]$.
\item To obtain existence also for negative times, one can apply
Theorem~\ref{thm1} to the time reversed current
\begin{equation}\label{jbar}
\bar\jmath(t,q) = \bigl( j^0(-t,q), -\sj(-t,q) \bigr)\,.
\end{equation}
The integral curves of $\bar\jmath$ are the time reverses of the
integral curves of $j$. Obviously, with $j$ also $\bar\jmath$
satisfies \eqref{current}. If $\bar\jmath$ satisfies \eqref{ass1},
 \eqref{ass3}, and, if appropriate, \eqref{ass4} for
  $T>0$, we obtain almost-sure global existence of $Q_q(t)$ on $[-T,0]$.
\item It suffices to consider in (\ref{ass1}), (\ref{ass3}) and (\ref{ass4})
for the sets $B$ instead of all bounded Borel sets just the balls
around the origin. This is because enlarging $B$ cannot shrink the
integral. For the same reason,  it suffices to integrate over
$\conf\setminus \Nodes_t$ instead of the not easily accessible sets
$\flow_t(B) \setminus \{\cemetery\}$.
\item Actually the proof of Theorem~\ref{thm1} works in the same way with the following slightly weaker
conditions. Instead of \eqref{ass3} it suffices to assume that
\[
  \forall B \in
    \Bdd(\conf)\,\,\,\exists\,R<\infty\,:\quad
    \int\limits_0^T dt \int\limits_{\flow_t(B) \setminus
     \{\cemetery\}
    } \hspace{-1.2em} dq \:\:\: {\bf 1}(|q|>R)\,\left| \sj (t,q)\cdot\frac{q}{|q|}\right| < \infty
    \,,
  \]
and \eqref{ass4} can be replaced by
\[
\forall B \in
    \Bdd(\conf)\,\,\,\exists\delta>0:\quad
\int\limits_0^T dt \int\limits_{\flow_t(B) \setminus \{\cemetery\}
    }\hspace{-1.2em} dq \,{\bf 1}\big(q\in (\Sing_\ell+\delta)\big)\, \frac{\left| \sj(t,q)\cdot e_\ell(q)\right|}{{\rm
    dist}(q,\Sing_\ell)}<\infty\,.
\]
We chose to state the theorem with the stronger assumption to
simplify the presentation and because the weaker assumptions will
not be used in the following.
\end{enumerate}

\begin{proof}[ of Theorem~\ref{thm1}]
Let $Q_q(t)$ be the maximal solution of \eqref{dQdtj} starting in
$q$, as described in Section~\ref{sec:CVF}. Since we deal only with
positive times in the following, we write $\tau_q$ for $\tau_q^+$.

There are three ways in which  $Q_q(t)$ can fail to exist globally:
the trajectory can approach a node, approach a point on the singular
set $\cup \Sing_\ell$, or escape to infinity in finite time.  More
precisely, {\it if $q\notin\Nodes_0$ and $\tau_q<\infty$ there
exists an increasing sequence $(t_n)_{n\in \NNN}$ with $t_n\to
\tau_q$ such that either there is $x\in
\Nodes_{\tau_q}\cup\bigcup_{\ell=1}^m\Sing_\ell$ with $Q_q(t_n)\to
x$ or $|Q_q(t_n)|\to\infty$.}

To see this, suppose that $\tau_q<\infty$ and that such a sequence
did not exist. Then $Q_q:=\{(t,Q_q(t)):t\in(0,\tau_q)\}\subset
(\RRR\times \conf)\setminus\Nodes$ would remain bounded and bounded
away from the complement $(\RRR\times \cup \Sing_\ell)\cup\Nodes$.
Since $(\RRR\times \conf)\setminus\Nodes$ is open, there would be a
compact set $K\subset( \RRR\times \conf)\setminus\Nodes$ such that
$Q_q \subset K^\circ$, with $K^\circ$ the interior of $K$. However,
 the vector field $(1,\sj/j^0)$ is $C^1$ on $(\RRR\times \conf)\setminus\Nodes$ and thus uniformly Lipschitz on $K$. Therefore, all of its
maximal integral curves either exist  for all times or hit the
boundary of $K$, in contradiction to the hypotheses.

Let now $q\notin\Nodes_0$ and $\tau_q\leq T$. If there is
$x\in\Nodes_{\tau_q}$ and $(t_n)$ such that $Q_q(t_n)\to x$,
 then $j^0(t_n,Q_q(t_n))\to 0$. Hence,
the total variation of $t\mapsto\log j^0(t ,Q_q(t ))$ up to time $T$
diverges, i.e.,\ $L_q=\infty$ where
\begin{equation}\label{Lq}
  \Vari_q =
  \int\limits_0^{\min(T,\tau_q)} \!\! dt \: \Bigl| \frac{d}{dt} \log
  j^0 \bigl( t,Q_q(t) \bigr) \Bigr|\qquad \mbox{for}\,\,q\in\conf\setminus\Nodes_0\,.
\end{equation}
We now show that $L_\infty:=\{q \in \conf\setminus\Nodes_0: \Vari_q
= \infty \}$ is a $\measure_0$-null set. For this it suffices to
show that for any bounded  set $B \in \Bdd(\conf)$, $B \cap
L_\infty$ is a $\measure_0$-null set. For this in turn, it suffices
that the average of $\Vari_q$ over $B$ relative to the measure
$\measure_0$ be finite:
\[
  \int\limits_B dq \: j^0(0,q) \, \Vari_q = \int\limits_B dq \:
  j^0(0,q) \int\limits_0^{\min(T,\tau_q)} \!\! dt \, \frac{\bigl|
  \frac{d}{dt} j^0 \bigl(t,Q_q(t) \bigr) \bigr|}{j^0(t,Q_q(t))} =
\]
[the order of integration can be changed since the integrand is
nonnegative]
\begin{eqnarray*}
  &=& \int\limits_0^{T} dt \int\limits_B dq \: j^0(0,q) \,{\bf 1}( \tau_q>t) \,\frac{\bigl| \frac{d}{dt} j^0 \bigl( t,Q_q(t)
  \bigr) \bigr|}{j^0(t,Q_q(t) )}\\& =& \int\limits_0^{T} dt
  \int\limits_{\flow_t(B) \setminus \{\cemetery\} } \hspace{-1.2em}
  \rho_t(dq') \frac{\bigl| (\partial/\partial t + \frac{\sj}{j^0} \cdot
  \nabla_{q'}) j^0(t,q') \bigr|}{j^0(t,q')} =
\end{eqnarray*}
[by Lemma~\ref{Lrhoj0}]
\[
  = \int\limits_0^{T} dt \int\limits_{\flow_t(B) \setminus
  \{\cemetery\} } \hspace{-1.2em} dq' \: \bigl| (\partial/\partial t + {\textstyle \frac{\sj}{j^0}} \cdot
  \nabla_{q'}) j^0(t,q') \bigr|
  \stackrel{\eqref{ass1}}{<} \infty.
\]
This shows $\measure_0(L_\infty)=0$ and thus that the solution
$Q_q(t)$ of \eqref{dQdtj} $\measure_0$-almost surely does not
approach a node during $[0,T]$.

Now we consider the cases  that either
\[
\lim_{n\to\infty} |Q_q(t_n)| = \infty\,.
\]
or
\[
\exists \,x\in \cup_{\ell=1}^m \Sing_\ell :\quad Q_q(t_n)\to x\,.
\]
 Hence, for such initial
conditions either the total variation of $t\mapsto|Q_q(t)|$  is
infinite, i.e., $D_q=\infty$ where
\[
 D_q = \int\limits_0^{\min(T,\tau_q)} dt  \,
  \Bigl| \frac{d}{dt} |Q_q(t)|\Bigr| \qquad\mbox{for}\,\,q\in\conf\setminus\Nodes_0\,,
\]
or the total variation of $t\mapsto\log {\rm
dist}(Q_q(t),\Sing_\ell)$ restricted to  $\Sing_\ell+\delta$ is
infinite for some $\ell\in\{1,\ldots,m\}$ and any $\delta>0$, in
particular for the one in \eqref{ass4}, i.e., $V_{q,\ell}=\infty$
where
\[
V_{q,\ell} = \int\limits_0^{\min(T,\tau_q)} \!\! dt \,{\bf
1}\hspace{-2pt}\left(Q_q(t)\in (\Sing_\ell+\delta)\right)\: \Bigl|
\frac{d}{dt} \log
  {\rm
dist}(Q_q(t),\Sing_\ell)
\Bigr|\qquad\mbox{for}\,\,q\in\conf\setminus\Nodes_0\,.
\]
 Therefore it suffices to show that $D_\infty:=\{q \in \conf\setminus\Nodes_0: D_q = \infty
\}$ and $V_{\infty,\ell}:=\{q \in \conf\setminus\Nodes_0: V_{q,\ell}
= \infty \}$ are $\measure_0$-null sets, and for this we proceed  as
for $L_\infty$.

Let $B \in \Bdd(\conf)$. Then (\ref{dQdtj}), followed by exactly the
same arguments as for $L_q$, shows that local expectations of $D_q$
are finite, i.e.\
\[
  \int\limits_B dq \: j^0(0,q) \, D_q  = \int\limits_0^{T} dt \int\limits_{\flow_t(B) \setminus
  \{\cemetery\} } \hspace{-1.2em} dq' \, \left|\sj \bigl(
   t,q' \bigr)\cdot \frac{q'}{|q'|}\right|
  \stackrel{\eqref{ass3}}{<} \infty\,.
\]
Hence $\measure_0(D_\infty)=0$.  For local expectations of
$V_{q,\ell}$ we obtain, again with (\ref{dQdtj}) and
Lemma~\ref{Lrhoj0}
\begin{eqnarray*}
\int\limits_B dq \: j^0(0,q) \, V_{q,\ell} &=& \int\limits_B dq \:
  j^0(0,q) \int\limits_0^{\min(T,\tau_q)} \!\! dt \,{\bf
1}\hspace{-2pt}\left(Q_q(t)\in (\Sing_\ell+\delta)\right) \,\left|
  \frac{\dot Q_q(t)\cdot e_\ell(Q_q(t))}{{\rm dist} (Q_q(t),\Sing_\ell)}\right|
   \\&=& \int\limits_0^{T} dt \int\limits_{\flow_t(B) \setminus
  \{\cemetery\} } \hspace{-1.2em} dq' \,{\bf
1}\hspace{-2pt}\left(q'\in (\Sing_\ell+\delta)\right) \,\left|
  \frac{J(t,q')\cdot e_\ell(q')}{{\rm dist} (q',\Sing_\ell)}\right|\stackrel{\eqref{ass4}}{<} \infty\,.
\end{eqnarray*}
Hence also $\measure_0(V_{\infty,\ell})=0$,  concluding the
existence part of the theorem.

It remains to show equivariance. Since the probability of reaching
$\cemetery$ before time $T$ vanishes, we have $\rho_t(\conf) = 1$
for all $t \in [0,T]$. Since $\rho_t \leq \measure_t$ by
Lemma~\ref{Lrhoj0} and $\measure_t(\conf) = 1$ by \eqref{j0norm}, we
must have $\rho_t = \measure_t$, which is equivariance.
\end{proof}

 \noindent \textbf{Remarks.}
\begin{enumerate}
\addtocounter{enumi}{4}
\item A reasoning closely related to our method of proof is also applied in
\cite{bmex}, Remark~3.4.6. There, an expression analogous to
\eqref{ass1} is used to control the probability of reaching an
$\varepsilon$-neighborhood of $\Nodes$ before letting
$\varepsilon\to 0$.  Apart from the fact that the argument is
applied there only to the nodes and not to singularities and
infinity, it is also unnecessarily complicated, mainly because it
considers an $\varepsilon$-neighborhood instead of fully exploiting
the integral \eqref{Lq}.

\item The proof of equivariance was the only place where we used the
property \eqref{j0norm} of a current vector field. The existence
statement of Theorem~\ref{thm1} holds as well if $j$ satisfies
\eqref{current} except for \eqref{j0norm}; in particular, we may
allow $\measure_t (\conf) = \infty$.

\item Here is another equivariance result that does not use
\eqref{j0norm}: \textit{Let $\conf$ be a configuration space as in
\eqref{conf} and let $j$ satisfy \eqref{current} except for
\eqref{j0norm}. Suppose that almost-sure global existence holds in
both time directions, starting from any time. Then the family of
measures $\measure_t$  is equivariant on $\RRR$.}

To see this, note that for equivariance we need to show merely that
$\conf \setminus \flowim_t$ is a $\measure_t$-null set, or, in other
words, that for $\measure_t$-almost every $q \in \conf$ the integral
curve of $j$ starting in $(t,q)$ reaches back in time to time $0$.
But this is immediate from almost-sure global existence in the other
time direction, starting at time $t$.

Thus, if both $j$ and $\bar\jmath$ as defined in \eqref{jbar} and
their time translates satisfy \eqref{ass1}, \eqref{ass3}, and, if
appropriate, \eqref{ass4} for all $T>0$, we obtain equivariance
without \eqref{j0norm}.
\item\label{remark8}
Condition~\eqref{ass3} can be replaced by the condition
\begin{equation}
\mbox{the first order derivatives of $\sj$ are bounded
on}\,\,[0,T]\times \conf\,.
\end{equation}
To show this, we show that under this assumption every unbounded
solution $Q_q(t)$ with $\tau_q\leq T$ has $L_q=\infty$, with $L_q$
defined in \eqref{Lq}.

 To see this, first
note that
 the solutions of \eqref{dQdtj} are reparameterizations of the
integral curves of $j$. In more detail, let $\gamma_q(s) = \bigl(
\gamma^0_q(s), \sgamma_q(s) \bigr)$ be the unique maximal integral
curve to $j$,
\begin{equation}\label{gammadef}
  \frac{d\gamma_q(s)}{ds} = j(\gamma_q(s)) \,,
\end{equation}
starting in $(0,q)\in\RRR\times(\conf\setminus\Nodes_0)$ and defined
for $s\in(\sigma_q^-,\sigma_q^+)$.
Since $j^0>0$ outside nodes, $\gamma^0_q(s)$ is monotonically
increasing, and hence the map
\[
  s\mapsto t_q(s) = \gamma_q^0(s) = \int\limits_0^s d\tilde s \:
  \frac{d\gamma^0_q}{d\tilde s} = \int\limits_0^s d\tilde s \:
  j^0(\gamma_q(\tilde s))
\]
is invertible on its image $(\tau_q^-,\tau_q^+)$, where $\tau_q^\pm
= \lim_{s \to \sigma_q^\pm} t_q(s)$, with inverse $s_q(t)$. Since
\[
  \frac{d}{dt} \sgamma_q(s_q(t)) = \frac{ \sj \bigl( \gamma(s_q(t))
  \bigr) }{ j^0 \bigl( \gamma(s_q(t)) \bigr) }\,,
\]
$Q_q(t) = \Gamma_q(s_q(t))$ is the unique maximal solution of
\eqref{dQdtj} with $Q_q(0)=q$; it is defined for
$t\in(\tau_q^-,\tau_q^+)$.

Now suppose that $|Q_q(t_n)|\to\infty$ for some $t_n\to\tau_q^+$.
Then also $|\Gamma_q(s_q(t_n))|\to\infty$. Since the derivatives of
$\sj$ are bounded, there are constants $A$, $R>0$ such that
$|J(t,q)|\leq A|q|$ for all $t\in[0,T]$ and all $q\in \conf$ with
$|q|>R$. Since $d\Gamma_q/ds= J(\gamma_q(s))$, it follows that
$|\Gamma_q(s)|\leq \max (|\Gamma_q(0)|,R)\,\E^{As}$; thus, an
integral curve of $j$ cannot escape to spatial infinity in a finite
interval of the parameter $s$; in other words, $\sigma_q^+=\infty$.
But then
\[
\tau_q^+ = \int_0^\infty ds\,j^0(\gamma_q( s))<\infty
\]
implies the existence of an increasing sequence $(s_n)$ with
$s_n\to\infty$ such that $j^0(\gamma_q(s_n))\to 0$, and therefore
$L_q=\infty$.
\end{enumerate}

\section{Global Existence of Bohm--Dirac Theory}\label{sec:Dirac}

The Dirac Hamiltonian for $N$ particles is
\[
 H_{\rm D} = - \sum_{i=1}^N \I c\hbar \valpha_i \cdot \nabla_{\vq_i}   + V(\vq_1, \ldots,
 \vq_N)\,,
\]
where we assume a nonsingular $V\in C^\infty(\RRR^{3N}, \,{\rm
Herm}(\CCC^{4^N}))$. According to  \cite{Chernoff},   $H_{\rm D}$ is
essentially self-adjoint on $C_0^\infty(\RRR^{3N}, \CCC^{4^N})$ and
we denote by $H_{\rm D}$ the unique self-adjoint extension.

Since the Dirac matrices $\valpha$ have eigenvalues $\pm 1$, the
velocities in \eqref{BohmDirac} are bounded by $c$. Consequently,
the Dirac current \eqref{BohmDiracj} satisfies $| \sj | \leq c
\,\sqrt{N}\, j^0$.   This fact makes the proof of global existence
particularly simple, as expressed in the following corollary to
Theorem~\ref{thm1}.


\begin{corollary}\label{Cbddv}  Let $\conf=\RRR^d$ and let   $j=(j^0,\sj)$ be a current as
defined in \eqref{current}.
 Suppose  that there is a global bound on velocities, i.e., a constant $c>0$ such
  that $| \sj | \leq c \, j^0$.
\smallskip

  Then for $\mu_0$-almost all $q \in \RRR^d$, the solution of \eqref{dQdtj} starting at
  $Q(0)=q$ exists for all times, and the family of measures $\measure_t$ is equivariant.
\end{corollary}

\begin{proof}[ of Corollary~\ref{Cbddv}]
We show that assumptions \eqref{ass1} and \eqref{ass3} of
Theorem~\ref{thm1} are satisfied for any $T>0$. The key observation
is that due to the bound on velocities, bounded sets in
configuration space stay bounded under the flow. More explicitly,
for any bounded  set $B \in \Bdd(\RRR^d)$ contained in, say, the
ball $B_r$ of radius  $r>0$ around the origin, $\flow_t(B) \setminus
\{ \cemetery \}$ will be contained in $B_{r + ct}$ and thus in $B_{r
+ cT}$ provided $t \in [0,T]$. Now $\bigl|  (\partial_t  +
\frac{\sj}{j^0} \cdot
  \nabla_q ) j^0 \bigr|\leq \bigl|  \partial_t j^0\bigr| + c\bigl|
  \nabla_q  j^0  \bigr|$, and
 the functions  $|\sj|$, $\bigl|  \partial_t j^0\bigr|$, and $c\bigl|
  \nabla_q  j^0  \bigr|$ are continuous and therefore  bounded on the compact set
$[0,T] \times \overline{B_{r + cT}}$. Hence the integrals in
\eqref{ass1} and \eqref{ass3} are finite. This implies existence for
all positive times. For negative times apply the same argument to
the time-reversed current $\bar \jmath$, for which the same velocity
bound holds.
\end{proof}

Applying Corollary~\ref{Lrhoj0} to Bohm--Dirac theory, we obtain
global existence of Bohm--Dirac trajectories under very general
conditions.

\begin{theorem}
Let $V\in C^\infty(\RRR^{3N}, \,{\rm Herm}(\CCC^{4^N}))$ and
$\psi(t) = \E^{-\I tH_{\rm D}}\psi(0)$ with $\psi(0)\in
C^\infty(\RRR^{3N}, \CCC^{4^N})\cap L^2(\RRR^{3N}, \CCC^{4^N})$ and
$\|\psi(0)\|=1$.

Then the solution $Q_q(t) = (\vQ_1(t),\ldots,\vQ_N(t))$ of
\eqref{BohmDirac} with $Q_q(0)=q$ exists globally in time for almost
all $q\in\RRR^{3N}$ relative to the measure
$\measure_0(dq)=|\psi(0,q)|^2dq$, and the $|\psi(t)|^2$
distributions are equivariant.
\end{theorem}

\begin{proof}
According to \cite{Chernoff}, for $\psi(0)\in C^\infty_0(\RRR^{3N},
\CCC^{4^N})$ one has $\psi(t) \in C^\infty_0(\RRR^{3N}, \CCC^{4^N})$
and $\psi(t,q) \in C^\infty (\RRR\times\RRR^{3N}, \CCC^{4^N})$. But
then linearity and the finite propagation speed (Proposition~1.1 in
\cite{Chernoff}) imply that $\psi(t,q) \in C^\infty
(\RRR\times\RRR^{3N}, \CCC^{4^N})$ also for $\psi(0)\in
C^\infty(\RRR^{3N}, \CCC^{4^N})\cap L^2(\RRR^{3N}, \CCC^{4^N})$.
Hence, the Dirac current \eqref{BohmDiracj} satisfies
\eqref{current}. Since $| \sj | \leq c\,\sqrt{N} \, j^0$,
Corollary~\ref{Cbddv} implies the theorem.
\end{proof}

\begin{corollary}
 Let now $\conf=\RRR^d\setminus\cup_{\ell=1}^m\Sing_\ell$, where $\Sing_\ell$ is a hyperplane with codimension $\geq 2$ for $\ell=1,\ldots,
 m$, and
 let   $j=(j^0,\sj)$ be a current as
defined in \eqref{current}.
 Suppose  that there is a global bound $c$ on velocities, $| \sj | \leq c \, j^0$, and that $\sj$ and the first order derivatives of $j^0$ are bounded on bounded sets.
\smallskip

  Then for $\mu_0$-almost all $q \in \RRR^d$, the solution of \eqref{dQdtj} starting at
  $Q(0)=q$ exists for all times, and the family of measures $\measure_t$ is equivariant.
\end{corollary}

\begin{proof}
First note that $\RRR^d\setminus\conf$ is a Lebesgue-null set and
hence also a $\measure_0$-null set. For $q\in\conf$ we apply
Theorem~\ref{thm1}. The conditions~\eqref{ass1} and \eqref{ass3} of
Theorem~\ref{thm1} follow as in the proof of Corollary~\ref{Cbddv}
using the fact that $\sj$ and the derivatives of $j^0$ are locally
bounded.

To check \eqref{ass4}, let $d_\ell$ be the dimension of $\Sing_\ell$
and assume without loss of generality that $\Sing_\ell$ contains the
origin. Then with $|\sj|\leq C$ on $B_{r+cT}^d$, the ball of radius
$r+cT$ around the origin in $\RRR^{d}$, and $B_{r+cT}^{d}\subset
B_{r+cT}^{d_\ell}\times B_{r+cT}^{d-d_\ell}$ we find that
\[
 \int_0^T dt\, \int\limits_{B_{r+cT}^d
    } dq \, \frac{\left| \sj(t,q)\right|}{{\rm
    dist}(q,\Sing_\ell)} \leq  T  \int\limits_{B_{r+cT}^{d_\ell}
    } dx\,  \int\limits_{B_{r+cT}^{d-d_\ell}
    } dy \,\frac{C}{|y|}<\infty\,.
\]
\end{proof}

\section{Global Existence of Bohmian Mechanics}\label{sec:Schr}

We now apply Theorem~\ref{thm1} to Bohmian mechanics and consider
the abstract Hamiltonian
\begin{equation}\label{BohmHDef}
H_0 = -{\textstyle \frac{1}{2} }  \left(
\mathfrak{m}^{-\frac{1}{2}}\bigl(  \nabla_q - \I A(q)
\bigr)\right)^2 \,{\bf 1}_{\CCC^k} + V(q)\,,\qquad D(H_0) =
C_0^\infty(\conf,\CCC^k)\,,
\end{equation}
where, for the moment, $A\in H^1_{\rm loc}(\RRR^d,\RRR^d)$ and $V\in
L^2_{\rm loc}(\conf, {\rm Herm}(\CCC^k))$.  The mass matrix
$\mathfrak{m}= {\rm diag}(m_1,\ldots,m_d)$ has positive entries
$m_i>0$. These conditions assure that $H_0$ is well defined and
symmetric on $C_0^\infty(\conf,\CCC^k)$. Since $H_0$ commutes with
complex conjugation, $H_0$ has at least on self-adjoint extension.
We also assume that $\conf= \RRR^d\setminus\cup_{\ell=1}^m
\Sing_\ell$ where each $\Sing_\ell$ is a $(d-3)$-dimensional
hyperplane in $\RRR^d$. As to be explained in the example below, for
$d=3N$ the coincidence set of $N$ particles moving in $\RRR^3$ has
exactly this structure and therefore singular pair-potentials like
the Coulomb potential are included. In these abstract terms the
Bohmian equation of motion reads
\begin{equation}\label{BohmA}
  \frac{d Q}{dt}(t) = \mathfrak{m}^{-1}\Im \frac{\psi^* \, \bigl(
  \nabla_{q} - \I A  \bigr) \psi} {\psi^* \,
  \psi} (t,Q(t))\,.
\end{equation}

\begin{theorem}\label{BohmThm}
Let $H$ be a self-adjoint extension of $H_0$ as in \eqref{BohmHDef}
with domain $D(H)$. Suppose that  for some $\psi(0)\in D(H)$ with
$\|\psi(0)\|=1$ the solution $\psi(t)=\E^{-\I tH}\psi(0)$ of the
Schr\"odinger equation
 satisfies
\begin{enumerate}\renewcommand{\labelenumi}{{\rm(\roman{enumi})}}
\item $\psi\in C^2(\RRR\times \conf,\CCC^k)$,
\item for every $T>0$ there is a constant $C_T<\infty$ such that
\[
\int_{-T}^T dt\,\left(\,\|\,|\nabla \psi(t)|\,\|^2+\|\,|A\,
\psi(t)|\,\|^2+\|\,A\cdot\nabla \psi(t)\,\|^2\,\right)<C_T\,.
\]
\end{enumerate}
Then the solution  $Q_q(t)$ of \eqref{BohmA} with $Q_q(0)=q$ exists
globally in time for almost all $q\in\RRR^d$ relative to the measure
$\measure_0(dq)=|\psi(0,q)|^2dq$, and the $|\psi(t)|^2$
distributions are equivariant.
\end{theorem}

\noindent \textbf{Remark.}
\begin{enumerate}

\addtocounter{enumi}{8}
\item Note that condition (i) in Theorem~\ref{BohmThm} is typically
satisfied only if the potentials $A$ and $V$ are sufficiently smooth
on $\conf$, more than we required after \eqref{BohmHDef}. We decided
to state the condition in terms of $\psi$ since the exact type of
smoothness required for $A$ and $V$ depends on, among other factors,
the dimension $d$.
\end{enumerate}

\begin{proof}[ of Theorem~\ref{BohmThm}] First note that $\RRR^d\setminus\conf$ is a Lebesgue-null set and hence also a $\measure_0$-null set.
For $q\in\conf$ we apply Theorem~\ref{thm1}. According to
Section~\ref{sec:CVF} and by virtue of  (i), the Schr\"odinger
current
\[
j(t,q) = \left(\,  \psi^*(t,q)  \psi(t,q),\,\mathfrak{m}^{-1}{\rm
Im}\, \psi^*(t,q)  (\nabla_q - \I A(q))\,\psi(t,q)\right)
\]
 satisfies \eqref{current}. We now check \eqref{ass1}, \eqref{ass3}
 and \eqref{ass4}, in order to prove existence for positive times.
 For negative times one concludes analogously by applying exactly the same arguments to the time reversed
 current.

With $\psi(t)=\E^{-\I tH}\psi(0)$, the Cauchy--Schwarz inequality,
and (ii) we obtain
\begin{eqnarray*}
\int_\conf dq \left| \partial_t j^0(t,q) \right| &=&
 \int_{\RRR^d} dq \left| \partial_t   \psi^*(t,q)
\psi(t,q)\right| \leq 2 \int_{\RRR^d} dq \left|  \psi^*(t,q)  H
\psi(t,q)  \right| \\ &\leq& 2\| H\psi(t)\| = 2\|H\psi(0)\|\,.
\end{eqnarray*}

For the second term in \eqref{ass1} we find, after a straightforward
computation involving Cauchy--Schwarz first on $\CCC^k$ and then on
$L^2(\RRR^d)$ and finally on $L^2([0,T])$, that
\begin{eqnarray*}
\int\limits_0^T dt\,\int\limits_{\conf\setminus\Nodes_t}
\hspace{-5pt}dq \left| { \frac{\sj}{j^0}} \cdot
  \nabla j^0(t,q) \right| &\leq &\frac{1}{m_0}\int\limits_0^T dt\,\int\limits_{\RRR^d} dq \left(
  |\nabla\psi(t,q)|^2 + |\psi(t,q)| \,|A(q)\cdot \nabla
  \psi(t,q)|\right)\\&\leq& \frac{ C_T+ \sqrt{TC_T}}{m_0}\,,
\end{eqnarray*}
 where   $m_0=\min\{m_1,\ldots,m_d\}$.
Hence, \eqref{ass1} holds. Analogously \eqref{ass3} follows  from
\[
\int\limits_0^T dt\,\int\limits_\conf
     dq\, | \sj (t,q)| \leq \int\limits_0^T dt\,\frac{1}{m_0}\int\limits_{\RRR^d}
     dq\, |\psi(t,q)|\, ( |\nabla \psi(t,q)| + |A(q)\psi(t,q)|)\leq
     \frac{\sqrt{T C_T}}{m_0}\,.
\]

We now come to \eqref{ass4}.  Since $\Sing_\ell$ is a
$(d-3)$-dimensional hyperplane, it can be written as
$\Sing_\ell=\{q\in\RRR^d:y_\ell(q) = a_\ell\}$ with $y_\ell:
\RRR^d\to\RRR^3$, $q\mapsto (q\cdot y_\ell^1, q\cdot y_\ell^2,q\cdot
y_\ell^3)$ where $y_\ell^1$, $y_\ell^2$, $y_\ell^3$ are 3 orthogonal
unit vectors normal to the hyperplane $\Sing_\ell$ and
$a_\ell\in\RRR^3$ a constant. The distance to the hyperplane is
given by ${\rm dist}(q,\Sing_\ell) = |y_\ell(q)-a_\ell|$.

To prove \eqref{ass4}  for $\delta=\infty$, we use the generalized
Hardy inequality introduced in \cite{bmex}, Equation~(25). It states
that for all $\phi\in H^1(\RRR^d,\CCC)$, the first Sobolev space,
\[
\int_{\RRR^d} dq \,\frac{|\phi(q)|^2}{4|y_\ell(q) - a_\ell|^2}\leq
\int_{\RRR^d} dq\,|\nabla\phi(q)|^2\,.
\]
Hence,
\begin{eqnarray*}\lefteqn{\hspace{-15pt}
\int\limits_0^T dt\,\int_\conf dq\, \frac{|J(t,q)\cdot
e_\ell(q)|}{{\rm dist}(q,\Sing_\ell)}  \leq
 \frac{1}{m_0}\int\limits_0^T dt\int_{\RRR^d}dq\, \frac{|
\psi^*(t,q) (\nabla - \I
A(q))\,\psi(t,q) |}{|y_\ell(q)-a_\ell|}}\\
&\leq& \frac{1}{m_0} \int\limits_0^T dt\int_{\RRR^d}dq\,
\frac{|\psi(t,q)| (|\nabla \psi(t,q)|+  |
A(q)\psi(t,q)|)}{|y_\ell(q)-a_\ell|}\\
&\leq& \frac{1}{m_0}\int\limits_0^T dt\left( \int_{\RRR^d}dq\,
\frac{|\psi(t,q)|^2}{|y_\ell(q)-a_\ell|^2}\right)^\frac{1}{2} \,
\left(\|\,|\nabla\psi(t)|\,\| + \|\,|A\psi(t)|\,\|\right)\\
&\leq &\frac{1}{m_0}\int\limits_0^T dt  \,
(2\|\,|\nabla\psi(t)|\,\|^2 +
\|\,|\nabla\psi(t)|\,\|\,\|\,|A\psi(t)|\,\|)\leq \frac{3C_T}{m_0}\,.
\end{eqnarray*}
\end{proof}

We shall not try to verify the assumptions of Theorem~\ref{BohmThm}
under as general as possible conditions on $A$ and $V$. Instead we
consider two examples where they can be checked without too much
effort.

Our first example concerns a molecular system in external fields.
More precisely we consider $N$ electrons  in $\RRR^3$ with
configuration $q=(\vq_1,\ldots,\vq_N)\in \RRR^{3N}$ interacting
through Coulomb potentials
\[
V_{\rm el}(q) =  \sum_{i=1}^{N-1} \sum_{j=i+1}^N \frac{1}{|\vq_i -
\vq_j|}
\]
in the electric  potential
\[
V_{\rm nu} (q) = -\sum_{i=1}^N \sum_{j=1}^M \frac{Z_j}{|\vq_i -
\vz_j|}
\]
of $M$ static nuclei located at $\vz_j\in\RRR^3$ with charges $Z_j$,
$j=1,\ldots,M$. Furthermore we allow for an external magnetic field
$\vB(\vx)= \nabla \times \vA(\vx)$ with $\vA\in
C^\infty(\RRR^3,\RRR^3)$ such that $\nabla\cdot \vA = 0$ and $\vB$
and $\vA$ are bounded. The Hamiltonian of the system thus is
\begin{equation}\label{Hatomic}
H_{\rm mol} =  \left(-{\textstyle \frac{1}{2} }  \sum_{i=1}^N\bigl(
\nabla_{\vq_i} + \I \vA(\vq_i) \bigr)^2  + V_{\rm el}(q)  + V_{\rm
nu} (q)\right){\bf 1}_{(\CCC^2)^{\otimes N}} - \sum_{i=1}^N
\vB(\vq_i)\cdot \vsigma_i
\end{equation}
with domain $D(H_{\rm mol}) = H^2(\RRR^{3N}, (\CCC^2)^{\otimes N})$.
Here $\vsigma_i$ is the vector of Pauli matrices acting on the spin
index of particle $i$. It is well known that $V_{\rm el}$, $V_{\rm
nu}$, and $\nabla_q$ are infinitesimally bounded with respect to
$\Delta_q$. Hence
 $H_{\rm mol}= -\frac{1}{2}\Delta_q + R$ with
\[
R:=\left(-{\textstyle \frac{1}{2} }  \sum_{i=1}^N\bigl(  2\I
\vA(\vq_i)\cdot \nabla_{\vq_i} - \vA(\vq_i)^2 \bigr)   + V_{\rm
el}(q) + V_{\rm nu} (q)\right){\bf 1}_{(\CCC^2)^{\otimes N}} -
\sum_{i=1}^N \vB(\vq_i)\cdot \vsigma_i
\]
 is self-adjoint by virtue of Kato's theorem.

\begin{corollary}\label{MolCor}
 Let $\psi(t) =\E^{-\I
tH_{\rm mol}} \psi(0)$ with $\psi(0) \in C^\infty(H_{\rm mol}) =
\cap_{n=1}^\infty D((H_{\rm mol})^n)$ and $\|\psi(0)\|=1$. Then  the
Bohmian trajectories $Q_q(t)$ exist globally in time for almost all
$q\in\RRR^{3N}$ relative to the measure $|\psi(0,q)|^2dq$, and the
$|\psi(t)|^2$ distributions are equivariant.
\end{corollary}

\begin{proof}
First note that $H_{\rm mol}$ is of the form \eqref{BohmHDef} with
$d=3N$ and $k=2^N$. The configuration space of the system is
\[
\conf=\RRR^{3N}\setminus \left(\left( \cup_{i=1}^{N-1}\cup_{j=i}^{N}
\{q\in\RRR^{3N}\hspace{-3pt}:\vq_i=\vq_j\} \right)\,\cup \,\left(
\cup_{i=1}^{N}\cup_{j=1}^{M}
\{q\in\RRR^{3N}\hspace{-3pt}:\vq_i=\vz_j\}\right)\right)\,,
\]
where  the $N  (N-1)/2$ electron--electron and the $N  M$
electron--nucleus coincidence  hyperplanes are all
$(3N-3)$-dimensional. As remarked above, $H_{\rm mol}$ is
self-adjoint on $H^2(\RRR^{3N}, (\CCC^2)^{\otimes N})$ and thus
satisfies the hypotheses of Theorem~\ref{BohmThm}. Hence it suffices
to check that $\psi(t)$ satisfies the
 hypotheses (i) and (ii) of Theorem~\ref{BohmThm}.  As for (i), note that all potentials in
\eqref{Hatomic} are $C^\infty$ on $\conf$. Then methods of elliptic
regularity can be applied to show that for $\psi(0)\in
C^\infty(H_{\rm mol})$ the solution of the Schr\"odinger equation
satisfies $\psi \in C^\infty(\RRR\times\conf)$. For details see the
appendix in \cite{bmex}. Finally notice that, since $\vA$ is assumed
to be bounded and since $\|\psi(t)\|=\|\psi(0)\|$, (ii) follows if
we can show that the kinetic energy $\|\,|\nabla\psi(t)|\,\|$
remains bounded. This is also standard but we give the short
argument anyway: since $R$ is infinitesimally bounded with respect
to $\Delta$, there are constants $0<a<1$ and $b>0$ such that
$\|R\phi\|\leq a\|\frac{1}{2}\Delta\phi\|+ b\|\phi\|$ for all
$\phi\in H^2=D(H_{\rm mol})$. Hence
\begin{eqnarray*}
\|\Delta\psi(t)\|&= &2\|({\textstyle\frac{1}{2}}\Delta -R +
R)\psi(t)\| \leq 2\|H\psi(t)\| + 2\|R\psi(t)\|\\&\leq&
2\|H\psi(t)\|+a\|\Delta\psi(t)\|+ 2b\|\psi(t)\|
\end{eqnarray*}
together with $\|H\psi(t)\|=\|H\psi(0)\|$ and
$\|\psi(t)\|=\|\psi(0)\|$ implies
\[
\|\Delta\psi(t)\| \leq \frac{2\|H\psi(0)\| + 2b\|\psi(0)\|}{1-a}=
C\,.
\]
But then also
\[
\|\,|\nabla\psi(t)|\,\|^2 = \langle \nabla\psi(t),\cdot
\nabla\psi(t)\rangle = -\langle  \psi(t),\Delta\psi(t)\rangle \leq
\|\psi(t)\|\,\|\Delta\psi(t)\|\leq \|\psi(0)\|\,C\,.
\]
\end{proof}

The last corollary coincides exactly with the result of \cite{bmex}
(see their Corollary~3.2).

\begin{corollary}
In \eqref{BohmHDef} let $k=1$,  $A=0$ and $V=V_1+V_2\in
C^\infty(\conf,\CCC)$, where $V_1$ is bounded below and $V_2$ is
$-\frac{1}{2}\Delta$-form bounded with relative bound $<1$. Then the
form sum $H= -\frac{1}{2}\Delta + V$ is a self-adjoint extension of
$H_0$ and for $\psi(t) = \E^{-\I tH}\psi(0)$ with $\psi(0) \in
C^\infty(H) = \cap_{n=1}^\infty D(H^n)$, $\|\psi(0)\|=1$, the
Bohmian trajectories $Q_q(t)$ exist globally in time for almost all
$q\in\RRR^d$ relative to the measure $|\psi(0,q)|^2dq$, and the
$|\psi(t)|^2$ distributions are equivariant.
\end{corollary}
\begin{proof}
For the statement about the form sum see \cite{Faris}. Again, as
shown in the appendix of \cite{bmex}, elliptic regularity implies
that $\psi\in C^\infty(\RRR\times\conf)$. Hence,  in order to apply
Theorem~\ref{BohmThm} it suffices to show that that  $\|\,|\nabla
\psi(t)|\,\|$ remains bounded. This follows by an argument analogous
to the one given in the proof of Corollary~\ref{MolCor}. For the
details see the proof of Corollary~3.2 in \cite{bmex}.
\end{proof}

\bigskip

\noindent {\bf Acknowledgements:} We thank Sheldon Goldstein for
helpful remarks and Florian Theil for pointing out to us reference
\cite{DiPernaLions}. R.T.\ thanks the Mathematics Institute of the
University of Warwick for hospitality and INFN for financial
support.


\begin{thebibliography}{28.}


\bibitem{BellHidden} Bell, J.\ S.: ``On the problem of hidden variables
   in quantum mechanics,'' Rev.\ Mod.\ Phys.\ \textbf{38}, 447--452
   (1966).  Reprinted in Bell, J.\ S.: \textit{Speakable and unspeakable
   in quantum mechanics}.  Cambridge: Cambridge University Press
   (1987), p.~1.




\bibitem{Karinthesis} Berndl, K.: \textit{Zur Existenz der Dynamik in
  Bohmschen Systemen}. Ph.~D.~thesis, Ludwig-Maximilians-Universit\"at
  M\"unchen. Aachen: Mainz Verlag (1995)


\bibitem{survey} Berndl, K., Daumer, M., D\"urr, D., Goldstein, S.,
  Zangh\`\i, N.: ``A Survey of Bohmian Mechanics,'' Il Nuovo Cimento
  \textbf{110B}, 737--750 (1995), and quant-ph/9504010


\bibitem{bmex} Berndl, K., D\"urr, D., Goldstein, S., Peruzzi, G.,
  Zangh{\`\i}, N.: ``On the global existence of Bohmian mechanics,''
  Commun.\ Math.\ Phys.\ \textbf{173}, 647--673 (1995), and
  quant-ph/9503013


\bibitem{Bohm52} Bohm, D.: ``A Suggested Interpretation of the Quantum
  Theory in Terms of ``Hidden'' Variables, I,'' Phys.\ Rev.\
  \textbf{85}, 166--179 (1952).  Bohm, D.: ``A Suggested Interpretation
  of the Quantum Theory in Terms of ``Hidden'' Variables, II,'' Phys.\
  Rev.\ \textbf{85}, 180--193 (1952)


\bibitem{Bohm53} Bohm, D.: ``Comments on an Article of Takabayasi
  concerning the Formulation of Quantum Mechanics with Classical
  Pictures,'' Progr.\ Theoret.\ Phys.\ \textbf{9}, 273--287 (1953)


\bibitem{BohmHiley} Bohm, D., Hiley, B.\ J.: \textit{The Undivided
    Universe: An Ontological Interpretation of Quantum Theory}. London:
  Routledge (1993)


\bibitem{Chernoff} Chernoff, P.\ R.: ``Essential Self-Adjointness of
Powers of Generators of Hyperbolic Equations,'' J.\ Funct.\ Anal.\
{\bf 12}, 401--414 (1973)

\bibitem{DiPernaLions} DiPerna, R.\ J., Lions, P.\ L.: ``Ordinary
differential equations, transport theory and Sobolev spaces,''
Invent.\ Math.\ {\bf 98}, 511--547 (1989)

\bibitem{DetlefBuch} D\"urr, D.: \textit{Bohmsche Mechanik als
    Grundlage der Quantenmechanik}. Berlin: Springer-Verlag (2001)




\bibitem{qe} D\"urr, D., Goldstein, S., Zangh\`\i, N.: ``Quantum
  Equilibrium and the Origin of Absolute Uncertainty,'' J. Statist.
  Phys. \textbf{67}, 843--907 (1992), and quant-ph/0308039






\bibitem{Faris} Faris, W.\ G.: \textit{Self-adjoint operators.} Lecture
Notes in Mathematics {\bf 433}. Berlin: Springer-Verlag (1975).

\bibitem{crex} Georgii, H.-O., Tumulka, R.: ``Global Existence of
  Bell's Time-Inhomogeneous Jump Process for Lattice Quantum Field
  Theory,'' to appear in Markov Proc. Rel. Fields
  (2004); math.PR/0312294 and mp\underline{\ }arc~04-11


\bibitem{schwerpunkt} Georgii, H.-O., Tumulka, R.: ``Some Jump
  Processes in Quantum Field Theory,'' to appear in the proceedings of
  the DFG Priority Program ``Interacting Stochastic Systems of High
  Complexity'', Springer-Verlag (2004), and math.PR/0312326


\bibitem{Holland} Holland, P.\ R.: \textit{The Quantum Theory of
    Motion}. Cambridge: Cambridge University Press (1993)

\bibitem{ode} Lefschetz, S.: \textit{Differential Equations:
Geometric Theory}. New York, London: Interscience Publishers (1957)



\end{thebibliography}
\end{document}